\documentclass[nofootinbib,preprint,amsmath,amssymb,floatfix,superscriptaddress]{revtex4}
\usepackage{graphicx}
\usepackage{dcolumn}
\usepackage{bm}

\begin{document}

\fontfamily{ptm}
\selectfont

\title{Emergence of Complex Dynamics in a Simple Model of \\ Signaling
Networks}

\author{Lu\'{\i}s A. N. Amaral}
\affiliation{\scriptsize
Department of Chemical and Biological Engineering,
Northwestern University, Evanston, IL 60208, USA}

\author{Albert D\'{\i}az-Guilera}
\affiliation{\scriptsize
Department of Chemical and Biological Engineering,
Northwestern University, Evanston, IL 60208, USA}
\affiliation{\scriptsize Cardiovascular Division,  Beth Israel Deaconess Medical
             Center, Harvard Medical School, Boston, MA 02215, USA}
\affiliation{\scriptsize Dept. F\'{\i}sica Fonamental,
         Universitat de Barcelona, 08028 Barcelona, Spain}

\author{Andre A. Moreira}
\affiliation{\scriptsize
Department of Chemical and Biological Engineering,
Northwestern University, Evanston, IL 60208, USA}

\author{Ary L. Goldberger}
\affiliation{\scriptsize Cardiovascular Division,  Beth Israel Deaconess Medical
             Center, Harvard Medical School, Boston, MA 02215, USA}

\author{Lewis A. Lipsitz}
\affiliation{\scriptsize Hebrew Rehabilitation Center for the Aged, Boston, MA 02131, USA}
\affiliation{\scriptsize Gerontology Division,  Beth Israel Deaconess Medical
             Center, Harvard Medical School, Boston, MA 02215, USA}

\maketitle

\bigskip
\noindent
Classification:
Applied Math.

\bigskip
\noindent
Corresponding author:
\begin{verse}
{
\addtolength{\baselineskip}{-.5\baselineskip}
Lu\'{\i}s A. Nunes Amaral  Ph.D. \\
Northwestern University
Department of Chemical and Biological Engineering \\
McCormick School of Engineering and Applied Science \\
2145 Sheridan Road \\
Evanston, IL 60208

Phone: ~~847/491-7850 \\
Fax:~~~~~ 847/491-7070\\
E-mail: amaral@northwestern.edu
}
\end{verse}

\bigskip
\noindent
Manuscript information: 14 text pages (including title pages,
references and 6 figures).

\bigskip
\noindent
Character count: $<$ 41,000 characters.

\bigskip

\noindent
Abbreviations: Cellular automata (CA), Detrended fluctuation analysis
(DFA), random Boolean networks (RBN)

\vfill
\break
%%%======================================================================
\bigskip

%\small
%\baselineskip=10pt
ABSTRACT

\bigskip

{\bf
\noindent
A variety of physical, social and biological systems
generate complex fluctuations with correlations across
multiple time scales. In physiologic systems, these
long-range correlations are altered with disease and aging.
Such correlated fluctuations in living systems have been
attributed to the interaction of multiple control systems;
however, the mechanisms underlying this behavior remain
unknown. Here, we show that a number of distinct classes of
dynamical behaviors, including correlated fluctuations
characterized by $1/f$-scaling of their power spectra, can
emerge in networks of simple signaling units.  We find that
under general conditions, complex dynamics can be generated
by systems fulfilling two requirements: i) a ``small-world''
topology and ii) the presence of noise. Our findings support
two notable conclusions: first, complex physiologic-like
signals can be modeled with a minimal set of components; and
second, systems fulfilling conditions (i) and (ii) are
robust to some degree of degradation, i.e., they will still
be able to generate $1/f$-dynamics.  }

\bigskip
\bigskip

\noindent
{\bf }

\vfill

\break

%\end{abstract}

%%%%%%%%%%%%%%%%%%%%%%%%%%%%%%%%%%%%%%%%%%%%%%%%%%%%%%%%%%%%%%
%%%%%%%%%%%%%% Introduction

\noindent
Complex systems are typically comprised of a number of interacting
units that communicate information and are able to process and
withstand a broad range of stresses
\cite{malik95,bassin94,goldberger02,buchman02}.  In physiology,
free-running healthy systems typically generate complex
output signals that have long-range correlations---i.e., a
$1/f$-decay of the power spectra for low
frequencies\footnote{ A power law decaying power spectrum
$S(f) \propto f^{-\beta}$ is the signature of a signal with
power-law decaying correlations. The case $\beta=2$
corresponds to a Brownian, while $\beta=0$ corresponds to a
completely uncorrelated ``white'' noise. The intermediate
case, $S(f)\propto 1/f$, is a ``compromise'' between the
small time-scale roughness but large time-scale smoothness
of white noise and the small time-scale smoothness but large
time-scale roughness of Brownian noise.}
\cite{peng95,ivanov99,amaral01}.  
Deviations from the 1/f pattern have been associated with
disease or aging on a number of
contexts~\cite{goldberger02,lipsitz02}.

In spite of its practical and fundamental interest
\cite{marshall00}, the origin of such correlated dynamics
remains an unsolved problem~\cite{buchman02}. Until
recently, attention has focused primarily on the complexity
of the specific physiologic sub-systems or on the nature of
the nonlinear interactions between
them~\cite{kauffman93,wolfram94,kaplan97}.  In particular,
Boolean variables, which can take one of two values, 0 or 1,
and Boolean functions have been extensively used to model
the state and dynamics of complex systems---see
\cite{kaplan97} for an introduction. The reason such a
``simplistic'' description may be appropriate arises from
the fact that Boolean variables provide good approximations
to the nonlinear functions encountered in many control
systems \cite{kauffman93,weng99,aldana02review,wolfram02}.
Random Boolean networks (RBNs) were proposed by Kauffman
\cite{kauffman93} as models of genetic regulatory networks,
and have also been studied in a number of other contexts
\cite{weng99,aldana02review}.  Wolfram
\cite{wolfram02}, in contrast, proposed that cellular automata (CA)
models---a class of ordered Boolean networks with {\it
identical\/} units---may explain the real-world's
complexity.  Neither of these two classes of models has been
shown to generate the complex dynamics with
$1/f$-fluctuations observed in healthy physiologic systems.

We propose here a new 
modeling approach (Fig. 1\textbf{a}) that departs from
traditional approaches in that we pay special attention to
the {\it topology\/} of the network of interactions
\cite{buchman02}, {\it and\/} the role of {\it noise\/} \cite{rao02}.
Our model is rooted in two considerations frequently
observed in real-world systems: (i) the units in the system
are connected mostly locally but also with some long-range
 connections, giving rise to so-called {\it small-world\/}
topology \cite{watts98,albert02}, and (ii) the interaction
between the units is affected by {\it noisy\/} communication
and/or by noisy stimuli
\cite{mar99,ozbudak02,elowitz02,blake03,isaacs03}. We
demonstrate that simple rules, such as the majority rule,
are able to generate signal with complex fluctuations under
simple but physiologically-relevant conditions.

%
%%%%%%%%%%%%%%%%%%%%%%%%%%%%%%%%%%%%%%%%%%%%%%%%%%% FIGURE 1
\begin{figure*}[t!]
 \includegraphics*[width=16cm]{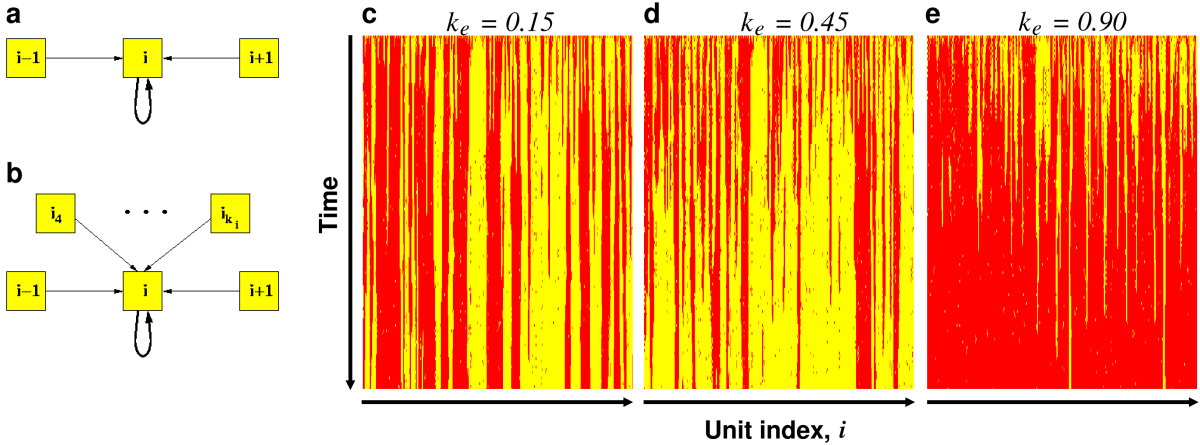}
 \label{f.model}
 \renewcommand\baselinestretch{1}
\caption{\footnotesize 
Emergence of complex dynamics in simple signaling networks.
Our model for signaling networks is defined as follows.
{\bf a,} The units comprising the network, which are located
on the nodes of a one-dimensional lattice, have
bi-directional nearest-neighbour connections.  {\bf b,} A
number $k_e N$ of additional unidirectional links is
established between pairs of randomly selected units, where
$k_e$ is the mean excess connectivity and $N$ is the number
of units in the system.  At time $t=0$, we assign to each
unit $i=1,\ldots,N$ a state $\sigma_i(0)$ randomly chosen
from the set $\{0,1\}$ and a Boolean function ${\cal F}_i$;
cf. Eq. (1).  This Boolean function (or rule) determines the
way the inputs are processed.  Each unit effectively
processes two inputs, one corresponding to the average state
of its neighbours and its own state.  With probability
$\eta$, a unit ``reads'' a random Boolean variable instead
of the state of a neighbour, where the parameter $\eta$
quantifies the intensity of the noise.  Note that the noise
does not alter the state of the units, just the value read
by its neighbour.  At each subsequent time step, each unit
updates its state synchronously according to its Boolean
function.
Time evolution of systems comprising 512 units with ${\cal
F}_i=232$ for all units and $\eta=0.1$, and \textbf{c} $k_e
= 0.15$, \textbf{d} $k_e = 0.45$, \textbf{e} $k_e = 0.90$.
A red dot indicates $\sigma_i(t) = 1$ while a yellow dot
indicates $\sigma_i(t) = 0$.  The three panels display the
time evolution for systems starting from the same initial
configuration and using the same sequence of random
numbers. Thus, the difference in the dynamics is due
uniquely to the different number of long distance links.
For $k_e = 0.15$ the system quickly evolves toward a
configuration with several clusters in which all the units
are in the same state.  The boundaries of these clusters
drift due to the noise but the state of the system $S(t)$ is
quite stable and the dynamics are close to Brownian noise.
In contrast, for $k_e = 0.90$ a large stable cluster
develops and the state of the system changes only when some
units change state due to the effect of the noise.  This
process yields white-noise dynamics.  For $k_e =0.45$
clusters are formed but they are no longer stable---unlike
what happens for small $k_e$.  In this case, information
propagates through the random links which can lead to a
change in the state of one or more units inside a cluster.
Our results suggest that because these long-range
connections exist on all length scales, they lead to
long-range correlations in the dynamics and the observed
$1/f$-behavior (Fig.~3\textbf{b}). }
\end{figure*}

%%%%%%%%%%%%%%%%%%%%%%%%%%%%%%%%%%%%%%%%%%%%%%%%%%%%%%%%%%%%%%%%%%%%%%%%%%%
%%%%%%%%%%%%%%%%%%%%%%%%%%%%% MODEL

\vspace{0.3cm}
{\noindent\bf Methods}

\vspace{0.2cm}
\noindent
{\it The model---}We place the Boolean units comprising the
network on the nodes of a one-dimensional ring and establish
bi-directional nearest-neighbor connections (Fig.~1).  Then,
we add $k_e N$ additional unidirectional links---where $k_e$
is the mean excess connectivity and $N$ is the number of
units in the system---between pairs of randomly selected
units.  Hence, each unit has a set of links through which
incoming signals arrive and that the unit then processes.
The state of each neighbour is replaced by a random value
with probability $\eta$---which parameterizes the intensity
of the noise.

We assign to each unit $i=1,\ldots,N$ a Boolean function ${\cal F}_i$,
which determines the way the states of the neighbors and its own state
are processed.  We restrict our study to Boolean functions that have
only two ``effective'' inputs: the state of the unit and the average
state of all other neighbours. This restriction yields 64 unique
symmetric Boolean functions (see Fig.~2 and the Supplementary Material) and
it has the advantage that it permits a topology-independent
implementation of the Boolean functions, thus enabling a systematic
study of the effect of different rules on the dynamics of the system.

%%%%%%%%%%%%%%%%%%%%%%%%%%%%%%%%%%%%%%%%%%%%%%%%%%%%%% FIGURE 2
\begin{figure}[t]
\begin{center}
   \includegraphics[width=8cm]{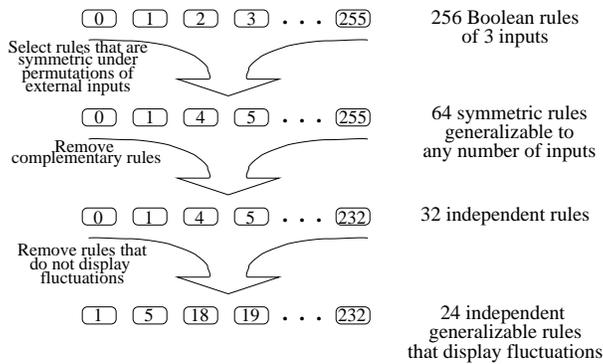}
 \renewcommand\baselinestretch{1}
\caption{\footnotesize 
Selection of Boolean rules for investigation.  Our goal is
to investigate Boolean functions that display non-trivial
dynamics and can be generalized to any number of inputs.  To
this end, we start from 256 rules of 3 inputs but then
restrict our attention to the ones that are symmetric under
permutations of the external inputs.  This selection results
in 64 Boolean rules. However, each rule has another that is
its complementary---i.e., display the same dynamics when
switching zeros and ones---or inverse---i.e., display the
same dynamics when taken every other step. Since these pairs
of rules have equivalent dynamics, we need to investigate
only 32 independent rules. Of these, eight do not display
fluctuations, even in the presence of noise, resulting in 24
independent rules that could present complex
fluctuations. The phase-spaces of each of these 24 rules are
shown in the Supplementary Material.
}
   \label{f.chart}
\end{center}
\end{figure}
%%%%%%%%%%%%%%%%%%%%%%%%%%%%%%%%%%%%%%%%%%%%%%%%%%%%%%%%%%%%%%%%%

\vspace{0.2cm}
\noindent
{\it Quantification of the dynamical behavior of the system---}We
start all of our numerical simulations with a random initial
configuration and let the system evolve synchronously according to the
rules of the model.  We define the state $S(t)$ of the system as the
sum of the states $\sigma_i$ of all the Boolean units
\begin{equation}
S(t) = \sum_i \sigma_i(t) \,.
\label{e.state}
\end{equation}
We record $S(t)$ during the course of the simulation (see
Figs.~3\textbf{a-c}) and quantify the complexity of the time series
generated in terms of its auto-correlation function
\cite{goldberger02}.  We apply the detrended fluctuation analysis
(DFA) method \cite{peng95} which quantifies long-range
time-correlations in the dynamical output of a system by means of a
single scaling exponent $\alpha$ (Fig.~3\textbf{d}): Brownian noise
yields $\alpha = 1.5$, while uncorrelated white-noise yields $\alpha =
0.5$.  For many physiologic signals, one observes $\alpha \approx 1$,
corresponding to $1/f$-behavior, which can be seen as a ``trade-off"
between the two previous cases \cite{goldberger02}.

%%%%%%%%%%%%%%%%%%%%%%%%%%%%%%%%%%%%%%%%%%%%%%%%%%% FIGURE 3
%%%%%%%%%%%%%%%%%%%%%%%%%%%%%%%%%%%%%%%%%%%%%%%%%%%%%%%%%%
\begin{figure*}[t!]
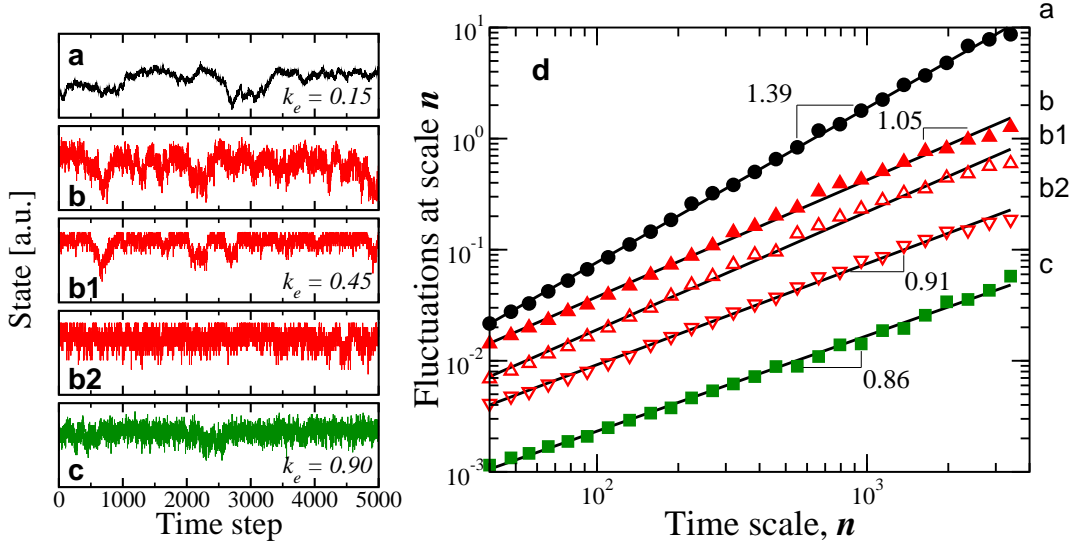

 \includegraphics*[width=5.25cm]{fig3-time_series}%%%%
~\includegraphics*[width=8.75cm]{fig3-dfa}
 \label{timedfa}
 \renewcommand\baselinestretch{1}
\caption{\footnotesize 
Quantification of the correlations in the state of Boolean
signaling networks.  As discussed in the text, we define the
state of the system as $S(t)=\sum \sigma_i(t)$.  We show
$S(t)$ for a system with $N=4096$ units, $\eta=0.1$, ${\cal
F}_i=232$, and $k_e = 0.90$, $k_e = 0.45$, $k_e = 0.15$.
The three values of $k_e$ lead to quite different dynamics
of the system.  {\bf a,} For a small number of random links,
the time correlations display trivial long-range
correlations such as found for Brownian noise. {\bf b,} For
an intermediate value of $k_e$, long-range correlations
emerge and the power spectrum displays a power-law behavior,
${\cal S}(f)\propto 1/f^\beta$, with $\beta \approx 1$.
Panels \textbf{b1-b2} display the state of the system
according to different definitions: In \textbf{b1} the state
of the system is defined as the sum of the states of a
random sample comprising 1/8th of all units, whereas in
\textbf{b2} the state of the system is defined as the sum of
the states of a block of contiguous units comprising 1/8th
of the systems.  Our results indicate that the evolution of
a subset of the population is similar to the dynamics of the
whole system.  {\bf c,} For a large number of random links,
$k_e=0.90$, the dynamics are less correlated.
\textbf{d,} 
Estimation of temporal auto-correlations of the state of the
system by the detrended fluctuation analysis (DFA) method
\protect\cite{peng95}. We show the log-log plot of the
fluctuations $F(n)$ in the state of the system, versus time
scale $n$ for the time series shown in
Figs.~3\textbf{a-c}. In such a plot a straight line
indicates a power-law dependence $F(n)\propto n^\alpha$.
The slope of the lines yields the scaling exponent $\alpha$,
which for a number of physiologic signals from free-running
healthy mature systems take values close to one
\protect\cite{goldberger02}.  The exponent $\alpha$ is
related to the exponent $\beta$ of the power spectrum of the
fluctuations, ${\cal S}(f)\propto 1/f^\beta$, through the
relation $\beta=2\alpha-1$. The data sets have been shifted
upward and the different sets correspond, from top to
bottom, to the time series shown in Panels 2\textbf{a-c}.  }
\end{figure*}

%%%%%%%%%%%%%%%%%%%%%%%%%%%%%%%%%%%%%%%%%%%%%%%%%%%%%%%%%%%%%%%%%%%%%%%%%%%
%%%%%%%%%%%%%%%%%%%%%%%%%%%%% RESULTS
\vspace{0.3cm}
{\noindent\bf Results}

\vspace{0.2cm}
\noindent
{\it Random Boolean networks---}The RBN model corresponds to a
completely random network with randomly selected Boolean rules for the
units.  As shown in Fig.~4\textbf{a}, we find white-noise dynamics for
essentially any pair of values of $k_e$ and $\eta$ within the ranges
considered, suggesting that {\it even in the presence of noise\/} a
system of random Boolean functions {\it cannot\/} generate
$1/f$-dynamics. This result is not unexpected, since the random
collection of Boolean functions comprising the system prevents the
development of any order or predictability in the dynamics.

\vspace{0.2cm}
\noindent
{\it CA models with small-world topology and noise---}We
systematically study the 64 symmetric rules (see
Supplementary Material and Fig.~2) for different pairs of values of
$k_e$ and $\eta$.  Some of the rules have parallels to
physiologically meaningful dynamics.  Rule 232 is a majority
rule, that is, each unit will be active next time-step only
if the majority of its neighbors is active now. Rule 50 is a
threshold rule with refractory time period, that is,
whenever the inputs of the neighbors surpass a certain
value a unit becames active in the next time-step and then will be
inactive for at least one time-step. The 64 symmetric rules
lead to three {\it qualitatively\/} distinct phase-spaces
(Figs.~4\textbf{b--d}).  Rule 232, the majority rule---which
is representative of the first type of
phase-space---displays three distinct types of dynamical
behavior (Fig.~4\textbf{b}): For small $k_e$, we find mostly
Brownian-like scaling.  For large $k_e$, we find mostly
white-noise dynamics.  Of greatest interest, for
intermediate values of $k_e$ and for a broad range of values
of the intensity of the noise $\eta$, we find
$1/f$-fluctuations.

Rule 50---which is representative of a second type of
phase-space---displays fewer types of dynamical behavior.  In
particular, we find only a narrow range of noise intensities (with a
weak dependence on $k_e$) for which the dynamics display
$1/f$-correlations.  For $\eta \stackrel{>}{\scriptstyle \sim} 0.1$,
the dynamics become uncorrelated (Fig.~4\textbf{c}).  Rule 160---which
is representative the third type of phase-space---displays white-noise
dynamics for all values of $k_e$ and $\eta$ (Fig~4\textbf{d}).

Note that for $k_e=0$---i.e., when the network is a one-dimensional
lattice---the model is not able to generate $1/f$-dynamics.  This
implies that in the context of the model, the existence of long-range
connections---i.e., the small-world topology achieved by making $k_e
\ne 0$---is an essential ingredient for the emergence of
$1/f$-dynamics.

%%%%%%%%%%%%%%%%%%%%%%%%%%%%%%%%%%%%%%%%%%%%%%%%%%% FIGURE 4
\begin{figure}[t]
\centerline{
  \includegraphics*[width=8cm]{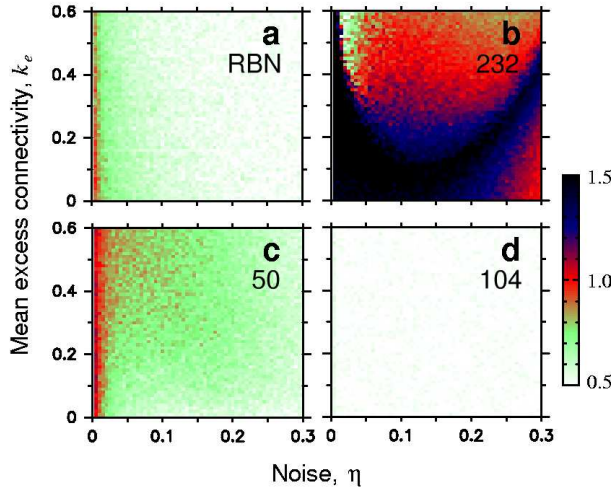}
}
  \renewcommand\baselinestretch{1}
 \caption{\footnotesize 
Systematic evaluation of the correlations in the dynamics
generated by different rules. We quantify the long-range
correlations in the dynamics by means of the DFA exponent
$\alpha$ \protect\cite{peng95} systematically estimated for
time-scales $40<n<4000$.  We show $\alpha$ for 3,721 pairs
of values of $k_e$ and the noise $\eta$ in the communication
between the units comprising the network. For all
simulations, we follow the time evolution of systems
comprising 4,096 units for a transient period lasting 8,192
time steps, and then record the time evolution of the system
for an additional 10,000 time steps.  In order to avoid
artifacts due to the fact that for some of the rules the
units switch states with period 2, we consider in our
analysis the state of the systems at every other time step.
  \textbf{a,} Random Boolean network (RBN) as defined by
  Kauffman \protect\cite{kauffman93}. Our results show that
  the dynamics generated by these systems are generally of
  the white-noise type with a weak dependence on the noise
  intensity and no dependence on the number of long-distance
  links.
  \textbf{b,} Rule 232, a.k.a. the majority rule.  This rule
  is representative of two other rules, numbers 19 and 1.
  Rule 232 displays a very rich phase-space with a variety
  of dynamical behaviors, all the way from white noise
  (represented in white and green) to Brownian noise
  (represented in black).
  \textbf{c,} Rule 50, a threshold rule with refractory
  period.  This rule is representative of eight other rules,
  numbers 5, 36, 37, 73, 77, 94, 108, 164.  These rules
  display a relatively simple phase-space with behaviors
  extending from white noise to $1/f$-noise.  The
  $1/f$-behavior is restricted to very small noise
  intensities and there is a very weak dependence on $k_e$.
  \textbf{c,} Rule 104.  This rule is representative of
  twelve other rules(see the Supplementary Material).  Their
  phase space is extremely simple, displaying only white
  noise behavior.
 }
  \label{f.normal}
\end{figure}

\vspace{0.3cm}
{\noindent\bf  Robustness of the findings}

\noindent
In order to determine the generality of the results presented above,
one needs to address the questions of how these findings are affected
by (i) changes in the topology of the network or (ii) ``errors'' in
the units' implementation of the rules.

Concerning (i), we note that the network topologies considered so far
span the cases of ordered one-dimensional lattices, small-world
networks, and random graphs \cite{amaral00}.  However, all networks
considered are comprised of units with approximately the same degree,
i.e., the same number of connections.  To investigate the role of the
distribution of number of connections, we also study networks which
span the range of empirically observed degree distribution: a
delta-distribution, an exponential distribution, and a power law
distribution.  The latter case corresponds to the so-called scale-free
networks \cite{albert02}.  Notably, we find that the picture of the
phase-space presented in Fig.~4{\bf b} does not get altered by these
changes in the degree distribution (see the Supplementary Material).

%%%%%%%%%%%%%%%%%%%%%%%%%%%%%%%%%%%%%%% FIGURE 5
%
\begin{figure}[t]
 \centerline{
  \includegraphics*[width=7.2cm]{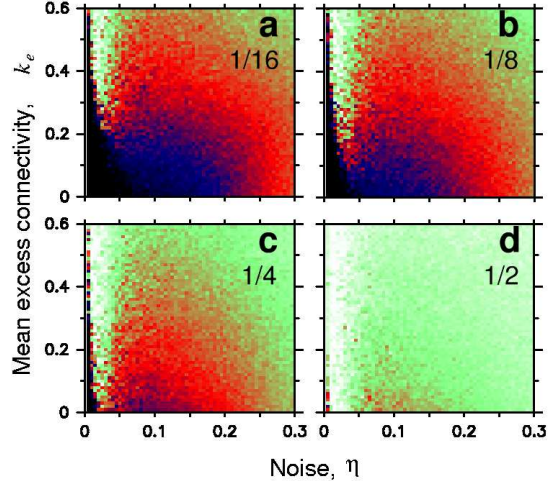}
}
 \label{mixing}
 \renewcommand\baselinestretch{1}
%
%\vspace*{-0.3cm}
  \caption{\footnotesize 
Phase-space for signaling networks with mixing of Boolean
rules.  We systematically calculate the exponent $\alpha$,
characterizing the correlations in the dynamics, for systems
composed of units operating according to rule 232 but with
some fixed fraction of units operating according to a
randomly selected symmetric Boolean rule.  Each value is an
average over 5 independent runs.
  \textbf{a,} 1/16th of the units operating according to a
  randomly selected rule.
  \textbf{b,} 1/8th of the units operating according to a
  randomly selected rule.
  \textbf{c,} 1/4th of the units operating according to a
  randomly selected rule.
  \textbf{d,} 1/2th of the units operating according to a
  randomly selected rule.
These figures suggest that the presence of random Boolean
functions leads to a decrease in the richness of the phase
space of the systems.  Specifically, if more than 1/4th of
all the units operate according to randomly selected Boolean
functions, then the phase-space displays mostly white-noise
dynamics.
  }
\end{figure}
%%%%%%%%%%%%%%%%%%%%%%%%%%%%%%%%%%%%%%%%%%%%%%%

In order to address (ii), we systematically explore the dynamical
behaviors in the phase-space defined by ($k_e,\eta$) for systems
composed of units operating according to either rule 232 or a randomly
selected rule.  Notably, we find that with as many as 1/4th of all
units operating according to random Boolean functions the model still
displays a rich phase-space including white, $1/f$ and Brownian noise
(Fig.~5).  This finding holds even if instead of using random Boolean
functions we consider a single Boolean function (Fig.~6).

%%%%%%%%%%%%%%%%%%%%%%%%%%%%%%%%%% FIGURE 6
%
\begin{figure}[t]
 \centerline{
  \includegraphics[width=7.2cm]{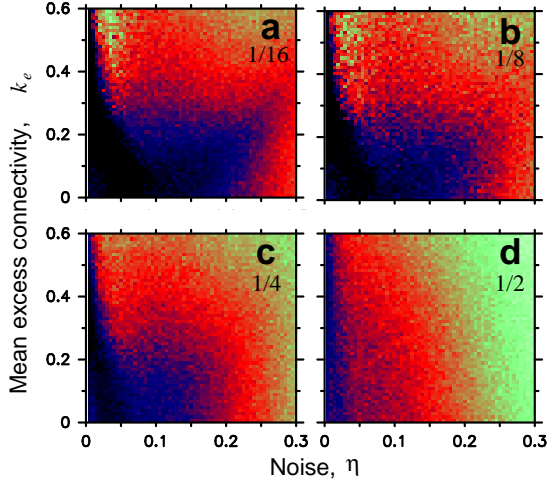}
}
 \renewcommand\baselinestretch{1}
%
%\vspace*{-0.3cm}
  \caption{\footnotesize 
Phase-space for signalling networks with mixing of two
Boolean rules.  We systematically calculate the exponent
$\alpha$, characterizing the correlations in the dynamics,
for systems composed of units operating according to either
rule 232 or rule 50.  Each value is an average over 5
independent runs.
\textbf{a,} 1/16th of the units operating according to
rule 50 and 15/16th operating according to rule
232. \textbf{c,} 1/8th of the units operating according to
rule 50 and 7/8th operating according to rule
232. \textbf{d,} 1/4th of the units operating according to
rule 50 and 3/4th operating according to rule
232. \textbf{e,} 1/2 of the units operating according to rule
50 and 1/2 operating according to rule 232.
When both rules are present in the system---and at least
50\% of the units operate according to rule 232---we still
find several distinct classes of dynamical behaviours,
including a wide range of parameter values that generate
$1/f$-noise.
  }
\end{figure}
%%%%%%%%%%%%%%%%%%%%%%%%%%%%%%%%%%%%%%%%%%%%%%%%%

%%%%%%%%%%%%%%%%%%%%%%%%%%%%%%%%%%%%%%%%%%%%%%%%%%%%%%%%%%%%%%%%%%%%%%%%%%%%
%%%%%%%%%%%%%%%%%% DISCUSSION

\vspace{0.3cm}
{\noindent\bf Discussion}

\noindent
Our results are notable for a number of reasons.
First, they demonstrate that a new model of signaling networks is
able to generate a broad range of behaviors reminiscent of those
observed in physiologic systems.  Second, we show for a rather
general class of models that a variety of dynamical behaviors can
only emerge under restrictive, but physiologically-relevant
assumptions; namely, the system must have a small-world topology, and
noise must be present.

An interesting aspect of our results is that some of the rules we
consider have plausible physiologic interpretation. For instance,
rule 232 is a rule in which a unit changes its state to that of the
majority of the incoming inputs.  A majority rule appears to be
operative in the central nervous system where multiple fibres
(excitatory or inhibitory) converge onto a single neuron.  Action
potentials converging in a neuron summate to bring the neuron to the
threshold for firing \cite{guyton00}.  A majority rule also appears to
be operative in the baroreflex control of the cardiovascular
system. The baroreflex is a feedback loop that continuously controls
heart rate by modulating the degree of sympathetic and parasympathetic
nervous system input to the sinus node of the heart. Changes in heart
rate values are determined by whichever input is dominant at the
moment \cite{guyton00}.

Our results demonstrates that complex fluctuations are
present even when a fraction of the units obeys randomly
chosen Boolean functions. This suggest that systems
comprised of units operating according to the majority will
be robust to the removal or failure of units. Additionally,
since this complex dynamics does not depend on a scale-free
topology, this system do not necessarily
displays the vulnerability to targeted attack observed in
scale-free networks \cite{albert00b}.  In biological systems
this robustness could support the ``physiologic reserve''
enabling an organism to overcome age- or disease-related
loss of system components.

Additionally, our findings raise the intriguing possibility that the
interactions within physiologic systems and their degradation with
aging and pathology may be symbolically mapped as a ``walk'' on the
phase space $(k_e,\eta)$.  According to this model, $1/f$-dynamics
similar to that found for healthy physiology are generated when noise
intensity and connectivity reside in a well-defined range
\cite{ivanov98}.  A loss of complexity with a breakdown of long-range
correlations would be anticipated when these parameters assume values
outside this range (Fig.~4). Support for this formulation comes from
analysis of heart rate dynamics with aging and disease, where
connectivity or coupling among system components is likely to be
degraded \cite{buchman02,goldberger02,lipsitz02}.  Similarly, evidence
suggests that decreased social connectedness, and the corresponding
decrease in ``noisy'' stimulation, may be associated with increased
cardiac mortality and decreased functional recovery from stroke or
dementia \cite{colantonio93,fratiglioni00}.

In a related way, the development of complex, adaptive
dynamics during the maturation of the organism may be
accounted for, at least in part, by the evolution of
appropriate connections (see
\cite{lipsitz97,yamamoto95} for empirical evidence).  Our model
predicts that the fraction of non-local connections has an
optimal range of values; hence, an excessive number of
certain types of inputs may also degrade functionality. Our
results are thus consistent with empirical evidence
suggesting that development and maintenance of healthy
function may require adjusting the number of connections.
Finally, the model may also provide a robust way to generate
fluctuations that closely resemble physiologic signals,
which could be implemented in medical devices such as
mechanical ventilators
\cite{suki98,boker02} and hormone infusion pumps
\cite{sturis95}.

{\small We thank A. Arenas, J. J. Collins, L. Glass, R. Guimera,
I. Henry, C.-C. Lo, G. Moody, J. M. Ottino, C.-K. Peng, C. J. Perez,
M. Sales-Pardo, H. E. Stanley, and G. Weisbuch for stimulating
discussions.  L.A.N.A. thanks a Searle Leadership Fund Award and a
NIH/NIGMS K-25 award.  A.L.G. thanks the support of NIH/NCRR (P41
RR13622) and of the Harold and Leila Y. Mathers Foundation. L.A.L.\ is
supported by NIH Grants AG04390 and AG08812; he also holds the Irving
and Edyth S. Usen and Family Chair in Geriatric Medicine at the Hebrew
Rehabilitation Center for Aged. A.D.-G.\ acknowledges financial
support from the Spanish Ministerio de Educaci\'{o}n, Cultura y
Deporte, DGES (Grant No.  BFM2000-0626 and BFM2003-08258) and European
Commission - Fet Open project COSIN IST-2001-33555.}

\end{document}